\documentstyle[prl,aps,epsf]{revtex}
\begin{document}

\title{ \bf Supercurrent Stability in a Quasi-1D Weakly Interacting Bose Gas}

\author{Yu.~Kagan, N.V.~Prokof'ev, and B.V.~Svistunov}

\address{Russian Research Center ``Kurchatov Institute," 123182 Moscow,
Russia }

\maketitle

\begin{abstract}
We discuss a possibility of observing superfluid phenomena in a
quasi-1D weakly interacting Bose gas at finite temperatures. 
The weakness of interaction in combination with
generic properties of 1D liquids can result in a situation when
relaxational time of supercurrent is essentially larger than the
time of experimental observation, and the behavior of the system
is indistinguishable from that of a genuine superfluid.
\end{abstract}

\bigskip


The discovery of Bose-Einstein condensation in ultra-cold dilute
gases \cite{discovery} has opened up a unique possibility for the study of 
fine aspects of superfluidity, including the specifics of
lower dimensionality. Primarily, the new opportunities come from
the availability of isolated systems, free of contact with walls
or any sort of substrate. A special interest is associated with
the prospects of creating one-dimensional systems. Actually,
one deals here with a quasi-1D situation, when the configuration of
a trap yields toroidal spatial particle distribution with small
transverse localization size $l_\perp$. At low temperature, and small
interaction energy per particle, $\epsilon_0$,
as compared to the transverse-motion level spacing, 
the system kinematics and collective dynamics are purely
one-dimensional, though the value of the effective interaction
is determined by 3D pair collisions (see below).

Strictly speaking, the 1D superfluidity cannot 
exist at $T \neq 0$. - At finite temperature the relaxational time
for supercurrent is finite, i.e., independent of the system size in the
limit $L \to \infty$.
Nevertheless, as will be shown below, analyzing the
situation with a dilute gas one finds that the relaxation rate 
of the supercurrent is {\it exponentially} reduced with
decreasing interaction, and that in a {\it quasi}-1D {\it
weakly} interacting system this can lead to a new quality when
at a finite temperature the lifetime of a supercurrent state
considerably exceeds the lifetime of the isolated gas sample.

We start with rendering basic notions and results for the low-energy
properties of 1D (super)fluids. First, we remind that in 1D there exists
a condition on superfluid density, $n_s$, and compressibility, $\kappa$,
determining whether the superfluid groundstate is stable
against either single impurity, or disorder, or an external
commensurate potential. This condition requires that the
dimensionless parameter ($m$ is the particle mass, $\hbar$=1)
\begin{equation}
K={1 \over \pi } \sqrt{m \over \kappa n_s} \equiv {cm \over \pi n_s }
\label{K}
\end{equation}
be less than corresponding critical value $K_c$ (here 
$c=\sqrt{n_s/m \kappa}$ is the sound velocity). For the
most important for our purposes cases of single impurity and
random potential the critical values of $K$ are $K_c=1$ \cite{Kane} and
$K_c=2/3$, respectively \cite{Schultz}. 
As we will see below, the case of quasi-1D weakly interacting
Bose gas corresponds to $K \ll 1$, so that the stability of the
superfluid groundstate is guaranteed.

The low-energy spectrum of any one-component 1D system in the superfluid phase
is exhausted by bosonic sound-like excitations and a collective state
supporting non-zero supercurrent, $J_I=2\pi n_s I/mL$, labeled by an 
integer index $I$ ~\cite{Haldane} (we consider a system of ring geometry)
\begin{equation}
H_0 = \sum_{q\ne 0} \omega_q b_q^{\dag } b_q 
+ \sum _{I} E_I d_I^{\dag } d_I \;,
\label{Ho}
\end{equation}
where $b_q^{\dag }$ creates a boson with energy $\omega_q = c q$ and
momentum $q=2\pi k/L$, where $k$ is integer. 
We have used a standard short-hand notation to specify the current state
 - obviously
$d_I^{\dag } d_I$ is either $1$ or $0$ and $\sum_I d_I^{\dag } d_I =1$. The 
energy of the current state is given by
\begin{equation}
E_I = {2\pi^2n_s \over mL }I^2 \;.
\label{EI}
\end{equation}
Note, that although the energy of the current state scales like $\sim 1/L$,
the momentum associated with it is system-size independent, 
$P_I=2\pi n_s I$. 

Within the long-wavelength effective Hamiltonian [describing the
superfluid state in terms of the phase field],  
$H = \int dx [ (n_s/2) (d \Phi /dx)^2 + (\kappa /2) \dot{\Phi
}^2 ]$,
where the original Bose field is written as $\Psi = \sqrt{n} e^{i\Phi }$,
bosonic excitations and current states (sometimes called ``zero
mode" states \cite{Haldane}) correspond to the following
decomposition
\[
\Phi = \phi (b) +2\pi I x/L\;, \;\;\;\;\;\oint (d \phi /dx) dx =0\;,
\]
where $x$ is a coordinate along the circle. Thus $I$ is nothing but a phase 
field 
circulation 
\begin{equation}
I={1 \over 2\pi } \oint (d \Phi /dx) dx \;.
\label{I}
\end{equation}

The very notion of the superfluid phase means that $I$ is a well-defined 
quantum number
even when rotational invariance is violated. In higher dimensions at 
$T<T_c$
relaxation times $\tau_I$ are extremely long with exponential dependence on 
the system
size, and for all relevant experimental time-scales, $\tau_{\rm exp}$, the 
condition
$\tau_{\rm exp} \ll \tau_I$ is satisfied. Special arrangements are to be made 
to 
shorten
$\tau_I$ (e.g., weak-link or tunnel-junction systems). To find relaxation 
times in 1D
we may use the transition Hamiltonian \cite{KPPS} which gives explicitly 
transition
amplitudes between states $\vert I, \{ N_q \} \rangle $ and 
$\vert I', \{ N_q'  \} \rangle $ for arbitrary rotational invariance 
breaking terms.
For the case of a local perturbation  $V(x)$ with spatial dimension smaller 
than
the correlation radius $1/mc$, which without loss of generality may be 
replaced by
\[
V(x) \to V_0 \, \delta(x-x_0) \; ,
\]
the transition Hamiltonian acquires a form 
\begin{equation}
H_{\mbox{\scriptsize int}} = g V_0 n \sum_{II'}  d_I^{\dag } d_{I'} e^{i2\pi 
n(I-I')x_0 }
\Lambda_{II'} (b) + h.c.  \;.
\label{Hint}
\end{equation}
Polaronic-type exponential operator $\Lambda (b)$ in this expression equals to
\begin{equation}
\Lambda_{II'} (b) = \exp \left\{ 
{i(I-I') \over \sqrt{K} } 
\sum_{q \ne 0} \left( 2\pi \over L\vert q \vert \right)^{1/2}
{\rm sgn} (q)(b_q-b_q^{\dag })e^{iqx_0} 
\right\}  \;.
\label{Lambda}
\end{equation}
The exact value of the numeric coefficient $g$ in Eq.~(\ref{Hint})
is outside the scope of the long-wavelength treatment used
to derive $H_{int}$ (it is of order unity for weakly-interacting  
Fermi gas).

Formally, the problem has reduced to the problem of particle
dynamics on a 1D lattice with Ohmic coupling to the oscillator
bath environment (see, e.g., review \cite{RMP}) after polaronic on-site
transformation. The dimensionless Ohmic coupling parameter
\begin{equation}
\alpha = (I-I')^2/K 
\label{alpha}
\end{equation}
is large in our case, $\alpha > 1$,
and ``particle''/current  dynamics is incoherent at any
finite temperature. We may then immediately utilize the well-known 
expression for the transition probability between states
$I$ and $I'$ which derives from the Golden-rule expression for
$H_{\mbox{\scriptsize int}}$ (see, e.g., Ref.~\onlinecite{KAGAN})
\begin{equation}
W=\vert gV_0 n \vert ^2 e^{-Z} {2\sqrt{\pi } \Omega \over \xi_{II'}^2 + 
\Omega^2 }
{\vert \Gamma [ 1+\alpha+i\xi_{II'}/ (2\pi T)] \vert^2 
\over \Gamma [1+\alpha] \Gamma [1/2+\alpha ] }
e^{\xi_{II'}/2T}    \;,
\label{W1}
\end{equation}
where
\begin{equation}
Z=2\alpha \ln \left(  {\gamma \epsilon_0 \over 2\pi T }  \right)   \;,
\label{z}
\end{equation}
\begin{equation}
\Omega = 2\pi \alpha T \;,
\label{Omega}
\end{equation}
\begin{equation}
\xi_{II'} = E_I-E_{I'} \;,
\label{xi}
\end{equation}
$\gamma $ is a numeric coefficient of order unity, and $\epsilon_0$
plays the role of the  high-energy cutoff
(at higher energies the dispersion curve is no longer sound-like and 
physics is determined by single-particle processes). 
Equation (\ref{W1}) was first derived in Refs.~\onlinecite{Grabert,Dorsey}.

To find the decay rate $\tau_I^{-1}$ one has to sum
Eq.~(\ref{W1}) over $I'$, but since we are interested in the
parameter range $\alpha \gg 1$ and $T \ll \epsilon_0$ the
dominant contribution comes from $I'=I \pm 1$. It follows then
from Eq.~(\ref{EI}) that $\xi \sim 4\pi^2n_sI/mL \ll T$ even for
$E_I \sim T$, and we may neglect current energy transfer to the
bosonic environment in Eq.~(\ref{W1}). Also, for $\alpha \gg 1$
we may write approximately $\Gamma [1+\alpha ]/\Gamma
[1/2+\alpha ] \approx \sqrt{\alpha }$. The final result may be
written then as
\begin{equation}
\tau^{-1}_I \sim  { \vert gV_0n \vert ^2  \over T\sqrt{\alpha }}
\left(  {2\pi T \over \gamma \epsilon_0}  \right)^{2\alpha }  \;.
\label{W2}
\end{equation}
We see that for large $\alpha$ the decay rate is severely
suppressed at low temperatures $T < T_0 \sim \epsilon_0/2\pi $.
Physically, this effect reflects small overlap between different
vacuum states corresponding to quantum numbers $I$ and $I'$.  In
practice, for $\alpha \gg 1$ there is a ``kinetic crossover"
between the superfluid and normal behavior, i.e., between frozen
and fast current relaxation, at $T \sim T_0$.

To estimate realistic parameters for ultra-cold atomic gases we have to start
from the original quasi-1D geometry. Let the transverse motion is confined by 
the 
symmetric parabolic potential with frequency $\omega_{\perp}$. Assuming that 
the 3D
scattering length $a$ is small as compared with the oscillator length 
$l_{\perp} =
1/\sqrt{m \omega_\perp }$, we may derive the potential energy per particle 
from the 3D relation 
\begin{equation}
\epsilon_0 = n {4\pi a \over m} \int d {\bf r_\perp} \,
\varphi_0^4(r_\perp) \equiv nU_{\rm eff} \; ,
\label{e0}
\end{equation}
where $\varphi_0$ is the wavefunction of the transverse motion
(here $n$ is the 1D particle density).
For the parabolic potential we have
\begin{equation}
\epsilon_0 = n2a \omega_\perp  \;.
\label{e00}
\end{equation}
Purely 1D kinematics requires two conditions to be satisfied
\begin{equation}
T\ll \omega_\perp\;, \;\;\;\;\;\;\;\;\;\;\;\;\;\; 
\epsilon_0 \ll \omega_\perp \;.
\label{cond}
\end{equation}
The second condition requires that a specific gas parameter,
$na$, be small: $na \ll 1$.
When deriving Eq.~(\ref{W2}) we assumed that $T \ll \epsilon_0$
to guarantee that transition rates between the supercurrent states
are negligible.
Obviously, this is a more severe restriction on the temperature
range than the first inequality in Eq.~(\ref{cond}). 

We now turn to the parameter $K$ for a translationally invariant system.
(Weak disorder does not change the value of $K$ dramatically.)
For the weakly interacting gas compressibility is given simply by inverse
$U_{\rm eff}$, and, in the low-energy limit, $n_s=n$. 
Thus,
\begin{equation}
K={1 \over \pi} \sqrt{ {mU_{\rm eff} \over n} }=
  {\sqrt{2}\over \pi} \left( {a  \over l_\perp } \right)^{1/2}
{1  \over (nl_\perp )^{1/2} } \;.
\label{K2}
\end{equation}
Experimentally it causes no difficulty to choose $nl_\perp > 1$, and to make
index $K$ very small or parameter $\alpha=1/K \gg 1$.

Let us consider an example of sodium gas assuming 1D density $n=
10^6$~cm$^{-1}$.
Then, $na\approx 0.25$ and 
$\epsilon_0 \approx  0.5 \times \omega_\perp $. For the
magnetic trap frequency $\omega_\perp = 3\times 10^4$~s$^{-1}$,
we find $n l_\perp \approx 30$, and extremely small $K < 0.01$ (!). 
The crossover temperature for the frozen current
dynamics is roughly $T_0 = \epsilon_0/2\pi \approx 2.5 \times
10^{-8}$~K, and within the experimental range. 

We note that the effect of supercurrent stability considered above
is essentially collective in nature and non-perturbative.
Our transition Hamiltonian is valid only if initial and final states
are given in terms of sound-like bosonic excitations - its validity
may be questioned if the energy change in the transition, $\xi$, exceeds
$\epsilon_0$. This imposes the  restriction that
interaction is not macroscopically small,
$U_{\rm eff} \gg  2\pi^2 /mL$. Also, even formally, for $\xi > \epsilon_0$
we find very fast relaxation rates from Eq.~(\ref{W1}) (for estimates one
may roughly replace in Eq.~(\ref{W2}) $2\pi T \to \xi $).
We would like thus to verify that with all the above parameters  we 
still may neglect energy (and momentum) transfer between the current and 
bosonic modes. Since $\xi =E_I-E_{I-1} = 4\pi^2n(I-1/2)/mL$, for $L =1$~cm
we find $\xi \approx 8 \times 10^{-9} (I-1/2)$~K, that is in the temperature
range between $4 \times 10^{-9}$~K and $T_0$ we still have $\xi <T$ for 
thermal
values of $I$ and system-size dependence is not crucial. On another hand, 
even if the energy has been dissipated into a single bosonic mode (in fact 
roughly
$\alpha \gg 1$ bosons are emitted/absorbed in the transition), the 
momentum of the boson would be only $ q = 4\pi (I-1/2)/KL $ and much smaller 
than
the momentum change of the current state $2\pi n$.

Thus, a quasi-1D interacting Bose gas with realistic
parameters can support a supercurrent at $T \neq 0$.
The stability of such a state is guaranteed by 
extremely large relaxational time as compared to
any reasonable experimental time. In connection with the
question of preparation of supercurrent state, it is
worthwhile to note that for an equilibrium 1D system at finite temperature
there is a finite probability to be found in a supercurrent
state. This naturally follows from the fact that
the energy of a supercurrent state scales like $1/L$, so
that at finite $T$ the statistical ensemble involves a large set of 
different numbers $I$. After a deep cooling, the system will
be found in a state with some particular $I$, randomly varying
from one experiment to another \cite{PS}.

This work was supported by the Russian Foundation for Basic
Research (Grant No. 98-02-16262) and by the Grant 
INTAS-97-0972 [of the European Community].


\begin{thebibliography}{99} \vspace{-0.8cm}
\bibitem{discovery} M.H. Anderson, J.R. Ensher, M.R. Matthews, C.E. Wieman, 
                    and E.A. Cornell, Science, {\bf 269}, 198 (1995).

\bibitem{Kane} Kane  C.L. and Fisher M.P.A.,
               Phys. Rev. Lett. {\bf 68}, 1220  (1992).

\bibitem{Schultz} Giamarchi T. and H.J.Schulz, 
                  Phys. Rev. B {\bf 37}, 325 (1988). 

\bibitem{Haldane} F.\ D.\ M.\ Haldane, Phys. Rev. Lett. {\bf 47}, 1840 (1981);
    J. Phys. C {\bf 14}, 2585 (1981).

\bibitem{KPPS} V.A.~Kashurnikov, A.I.~Podlivaev, N.V.~Prokof'ev,
 and B.V.~Svistunov, Phys. Rev. B, {\bf 53}, 13091 (1996).

\bibitem{RMP} A.J. Leggett, S. Chakravarty, A.T. Dorsey, M.P.A.
                   Fisher, A. Garg, and W. Zwerger,
                   Rev. Mod. Phys. {\bf 59}, 1 (1987).

\bibitem{KAGAN} Yu. Kagan and N.V. Prokof'ev, Zh. Eksp. Teor. Fiz., 
                {\bf 96}, 2209 (1989) [Sov. Phys. JETP - {\bf 69}, 1350];
                Yu. Kagan and N.V. Prokof'ev, in {\it Quantum Tunneling in 
                Condensed Matter}, eds. Yu. Kagan and A.J. Leggett,
                North-Holland, Elsevier, 37-143 (1992)

\bibitem{Grabert} H. Grabert and U. Weiss, 
                  Phys. Rev. Lett. {\bf 54}, 1605 (1985).

\bibitem{Dorsey} M.P.A. Fisher and A.T. Dorsey, 
                 Phys. Rev. Lett. {\bf 54}, 1609 (1985).

\bibitem{PS} N.V.~Prokof'ev and B.V.~Svistunov, submitted to Phys. Rev. B 
             (cond-mat/9907203).

\end{thebibliography}
\end{document}